\documentclass[
]{ceurart}

\usepackage{listings}
\usepackage{graphicx}
\usepackage{sepfootnotes}
\usepackage{csquotes}

\usepackage[dvipsnames,svgnames]{xcolor} 
\usepackage[normalem]{ulem} 

\makeatletter
\font\uwavefont=lasyb10 scaled 700
\def\spelling{\bgroup\markoverwith{\lower3.5\p@\hbox{\uwavefont\textcolor{Red}{\char58}}}\ULon}
\def\grammar{\bgroup\markoverwith{\lower3.5\p@\hbox{\uwavefont\textcolor{LimeGreen}{\char58}}}\ULon}
\def\phrasing{\bgroup\markoverwith{\lower3.5\p@\hbox{\uwavefont\textcolor{RoyalBlue}{\char58}}}\ULon}

\newcommand\remove{\bgroup\markoverwith{\textcolor{red}{\rule[0.5ex]{2pt}{0.4pt}}}\ULon}
\newcommand\insertion{\bgroup\markoverwith{\textcolor{Green}{\rule[-0.5ex]{2pt}{0.6pt}}}\ULon}
\makeatother

\begin{document}

\copyrightyear{2024}
\copyrightclause{Copyright for this paper by its authors.
  Use permitted under Creative Commons License Attribution 4.0
  International (CC BY 4.0).}

\conference{AMW 2024: 16th Alberto Mendelzon International Workshop on Foundations of Data Management, September 30th–October 4th, 2024, Mexico City, Mexico}

\title{Opportunities for Shape-Based Optimization of Link Traversal Queries}

\author[1]{Bryan-Elliott Tam}[%
email=bryanelliott.tam@ugent.be,
orcid=0000-0003-3467-9755
]
\cormark[1]

\author[1]{Ruben Taelman}[%
orcid=0000-0001-5118-256X,
email=ruben.taelman@ugent.be,
url=https://www.rubensworks.net,
]
\author[1]{Pieter Colpaert}[%
orcid=0000-0001-6917-2167,
email=pieter.colpaert@ugent.be,
url=https://pietercolpaert.be
]
\author[1]{Ruben Verborgh}[%
orcid=0000-0002-8596-222X,
email=ruben.verborgh@ugent.be,
url=https://ruben.verborgh.org
]

\cortext[1]{Corresponding author.}

\address[1]{IDLab,
Department of Electronics and Information Systems, Ghent University – imec}


\maketitle

\begin{abstract}
    Data on the web is naturally unindexed and decentralized.
    Centralizing web data, especially personal data, raises ethical and legal concerns.
    Yet, compared to centralized query approaches,
    decentralization-friendly alternatives such as Link Traversal Query Processing (LTQP)
    are significantly less performant and understood.
    The two main difficulties of LTQP are the lack of apriori information about data sources and the high number of HTTP requests.
    Exploring decentralized-friendly ways to document unindexed networks of data sources could lead to solutions to alleviate those difficulties.
    RDF data shapes are widely used to validate linked data documents, therefore, it is worthwhile to investigate their potential for LTQP optimization.
    In our work, we built an early version of a source selection algorithm for LTQP using RDF data shape mappings with linked data documents and measured its performance in a realistic setup.
    In this article, we present our algorithm and early results, thus, opening opportunities for further research for shape-based optimization of link traversal queries.
    Our initial experiments show that with little maintenance and work from the server, our method can reduce up to 80\% the execution time and 97\% the number of links traversed during realistic queries.
    Given our early results and the descriptive power of RDF data shapes it would be worthwhile to investigate non-heuristic-based query planning
    using RDF shapes. 
\end{abstract}

\section{Introduction}

\sepfootnotecontent{fn:typeIndex}{
  \href{https://solid.github.io/type-indexes/}{https://solid.github.io/type-indexes/}
}

The World Wide Web is a naturally decentralized database.
Centralizing large web segments in single endpoints provides easier query interfaces and faster query execution times.
However, data centralization can lead to practices that raise ethical and legal concerns, making the exploration of decentralization-friendly query paradigms a relevant research topic.
The query languages webSQL~\cite{Mendelzon1996} and \href{https://www.w3.org/TR/sparql11-query/}{SPARQL} propose mechanisms to capture decentralized web data with conjunctive queries.
However, webSQL relies on web indexing~\cite{Mendelzon1996}.
Indexing processes can be expensive, particularly on the scale of the web, and necessitate frequent updates, furthermore, they can be restrictive by excluding some sources thus hindering the natural serendipity of the web.
SPARQL solutions rely on the publication of linked data.
Linked data in their structure particularly with the presence of IRI gives the opportunity to find more related information without indexes.
However, most query processing over linked data is performed in centralized and federated setups, leaving indexing-independent approaches largely experimental.

Link Traversal Query Processing (LTQP)~\cite{Hartig2012} is a method to query unindexed networks of linked data documents.
The method consists of answering a query using an evolving triple store.
This evolving triple store is continuously updated with data acquired by the query engine through the recursive dereferencing of IRIs from the store.
The process is started with a set of IRIs provided by the user to the engine.
While LTQP enables live exploration of environments without prior indexing, it leads to some difficulties.
One of them is the pseudo-infinite search domain derived from the size of the World Wide Web~\cite{hartig2016walking}.
Additionally, HTTP requests can be very slow and unpredictable making their execution the bottleneck of the method~\cite{hartig2016walking}.
Reachability criteria~\cite{Hartig2012} are a partial answer to this problem by defining completeness based on the traversal of URIs
contained in the internal data source of the engine instead of on the acquisition of all the results or the traversal of the whole web.
Another difficulty is the lack of a priori information about the sources rendering query planning challenging.
To alleviate this problem, the current state-of-the-art consists of using carefully crafted heuristics for joins ordering~\cite{Hartig2011}.
The limitations of the heuristics approach are usually of little importance because the main bottleneck is the high number of HTTP requests.

Earlier LTQP research has focused on the open web.
More recently, LTQP research has shifted its focus to environments where the structure of data publication provides useful information for query optimization.
This line of research uses \emph{structural assumptions}~\cite{Taelman2023} to guide query engines~\cite{verborgh2020guided} towards relevant data sources.
Structural assumptions act as contracts between the data provider and the query engines stipulating that within a certain subdomain of the web, information meeting a specific constraint can be found.
The use of structural assumptions has been studied in Solid \cite{Taelman2023}.
The method involves the utilization of the 
\href{https://solidproject.org/TR/protocol#resources}{solid storage} hypermedia description~\cite{Fielding} to locate all the resources of a pod. 
This hypermedia description is not expressive enough to capture the content of the resources of a pod, thus, for query-aware optimizations, the \href{https://solid.github.io/type-indexes/}{type index specification}~\sepfootnote{fn:typeIndex} is additionally used.
The type index formulation proposes a more declarative approach~\cite{Taelman2017} by mapping RDF classes with sets of resources.
By using those structural assumptions it is possible to reduce the query execution time of realistic queries to the extent where the bottleneck is not the execution of HTTP requests but the suboptimal heuristic-based query plan~\cite{eschauzier_quweda_2023, Taelman2023}.
Yet, for multiple queries the high number of HTTP requests remains the main bottleneck~\cite{eschauzier_quweda_2023}.
It is reasonable to hypothesize that a significant portion of those HTTP requests lead to the dereferencing of documents containing data that do not contribute to the result of the query.
Hence, investigating more descriptive structural assumptions is a relevant research endeavor.

In this article, we propose to use RDF data shapes as the main mechanism for a structural assumption in the form of a shape index.
RDF data shapes are mostly used in data validation~\cite{Gayo2018a} thus, they provided a good formalization to describe the structure of data.
Additionally, to a lesser extent, they have been used for query optimizations~\cite{kashif2021}.
The shape index is an early effort for data summarisation of decentralized datasets~\cite{Stuckenschmidt2004,Goldman1997, Harth2010} within networks of unindexed linked data documents.
The current focus of the index is source selection.
However, we foresee opportunities to use a similar approach for link queue ordering and query planning.
This paper presents our preliminary work on data discovery and link pruning thus tackling the problem of the large search space of LTQP queries in linked data environments with structure.

\section{Shape Index and Query-Shape Containment}

\begin{figure}[h!]
    \centering
    \includegraphics[width=0.7\textwidth]{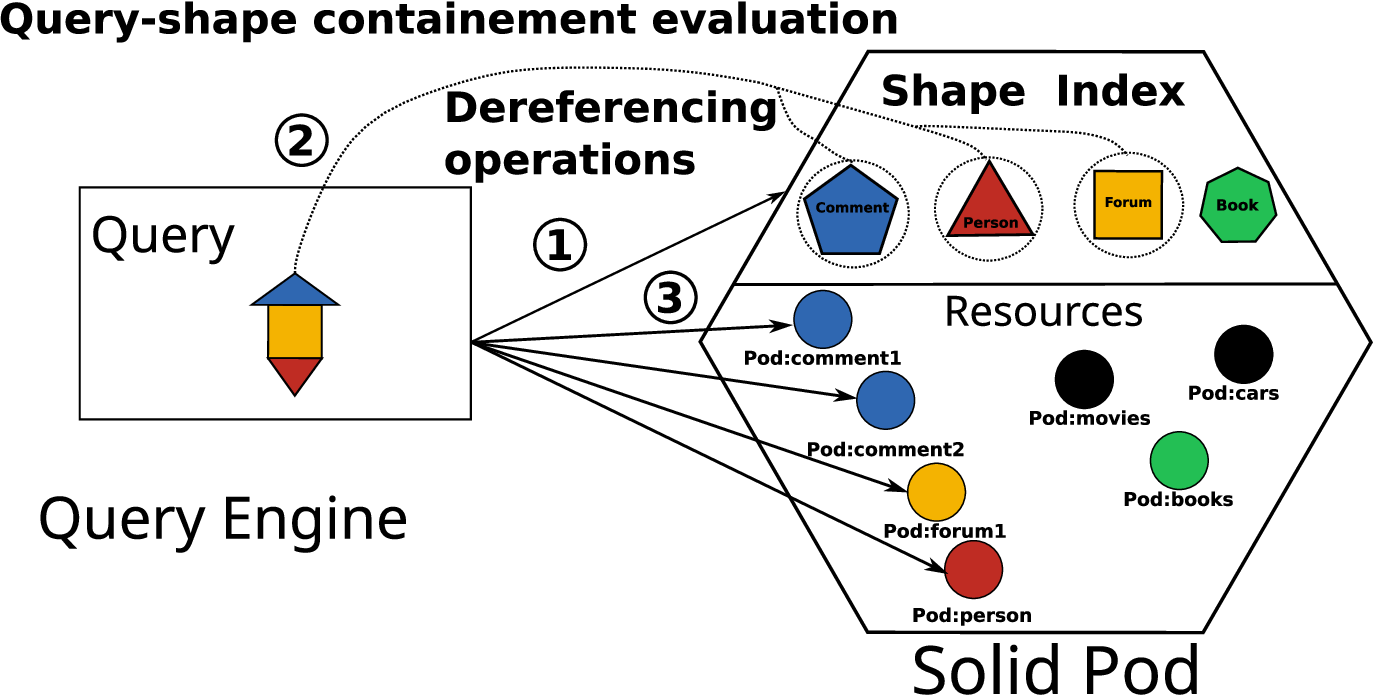}
    \caption{First, the shape index is dereferenced, 
    then the \emph{query-shape containment} operations are performed in the query engine and lastly, only the relevant resources are dereferenced.}
    \label{fig:shape_index}
\end{figure}

\sepfootnotecontent{fn:shapetrees}{
    \href{https://shapetrees.org/TR/specification/}{https://shapetrees.org/TR/specification/}
}

\sepfootnotecontent{fn:solidPrivacy}{
In this work, we do not take into consideration confidentiality restrictions.
}

\sepfootnotecontent{fn:litShapeComparaison}{
There exist comparisons between the shape and class definition approaches in the context of data validation~\cite{demeester_swj_2021} but it is left to be determined if their frame of comparison is compatible with our current problem
and foreseen opportunities.
}
\sepfootnotecontent{fn:domain}{
From a perspective where the domain is composed of URLs leading to linked data documents and the codomain is composed of the documents with their RDF content.
}

The RDF specification does not enforce schemas on data.
However, the data published often adhere to an implicit schema due to the nature of its modeled object and the formulation of RDF~\cite{Neumann2011CharacteristicSA}.
From those observations, implicit schemas have been used with success for query optimization~\cite{Neumann2011CharacteristicSA, Meimaris2018HierarchicalCS}.
We propose to adapt those methods for LTQP by using explicit data schemas provided by the data provider in our source selection process.

\subsection{Shape Index}

Our method introduces the concept of a \emph{shape index} to reduce query execution time by minimizing unnecessary dereferencing of RDF documents within web subdomains (sets of URLs or URL patterns).~\sepfootnote{fn:domain}
We define a shape index as a set of mappings between RDF data shapes and sets of resources.
This mapping concept shares similarities with shape mapping~\cite{Gayo2018} and target declarations~\cite{Gayo2018Shacl}.
However, instead of mapping shapes to RDF subgraphs, the shape index maps shapes to sets of documents.
The shape index also shares commonalities with \href{https://shapetrees.org/}{shape trees},~\sepfootnote{fn:shapetrees} however, it is designed to be a simpler formulation focused on query optimization.
The shape index has a range of applications defined in a domain and a flag indicating if the index is \emph{complete}.
A shape index is complete when every resource in the domain is associated with a shape within the shape index.
In a shape index when a shape is mapped to a set of RDF resources then the shape \emph{must} validate those resources.
Furthermore, every set of triples respecting the shape in the domain \emph{must} be located inside one resource of the set.

\subsection{Query-Shape Containment}

In order to determine before the traversal of a whole subdomain which resources are useful and which can be pruned, the query engine solves a \emph{query-shape containment} problem over the shape of the index analogous to the classic query containment problem~\cite{afariQCE, Spasi2023, Chekol2018}.
The query containment problem consists of determining independent of the data source if the results of a query will be a subset of the results of another query.
We propose to express RDF data shapes into \texttt{SELECT} SPARQL queries ($Q_{s}$)~\cite{delva2023, spapeExpressionConvert, labragayo2017validating, Corman2019} and apply similar resolution methods to query containment problems.
Due to the explicit domain definition of the index, this approach is adaptative, 
thus, the query engine can start its processing with permissive reachability criteria
such as $c_{all}$~\cite{Hartig2012} or the Solid state-of-the-art reachability criteria~\cite{Taelman2023}
and not suffer from the associated longer execution time during the traversal of environments containing a shape index.
The source selection process is schematized for a single (sub)domain in Figure~\ref{fig:shape_index}.
The process starts with the discovery of the shape index in the current (sub)domain.
In the case of Solid, the index can be at the root of the pod to be easily discoverable.~\sepfootnote{fn:solidPrivacy}
After the dereferencing of the index, the analysis is started inside the query engine.
The analysis consists of interpreting the binding results (homomorphism and "partial" homomorphism) of the \emph{query-shape containment} problem.
The algorithm divides the query from the user into multiple star patterns with their dependent star pattern ($Q_{star}$).
After the division of the query, the queries are pushdown~\cite{Stuckenschmidt2004, Yang2021FlexPushdownDBHP} to the level of source selection to evaluate if the $Q_{star}$ are contained inside the $Q_s$ of the shape index.
If all the $Q_{star}$ are contained in a $Q_{s}$ or have no binding with any $Q_{s}$
the reachability criterion is adapted to ignore all the resources not linked to a $Q_{s}$ even if the shape index is \emph{incomplete}.
If the shape index is \emph{complete} and not all the $Q_{star}$ are contained in a $Q_{s}$ the reachability criterion can be adapted
to visit every resource in relation to a $Q_{s}$ with a partial binding with a $Q_{star}$.
In a similar case with an \emph{incomplete} shape index, the query engine can only use the shape index for data discovery.
This case is similar to the usage of the type index but with a more reaching ability to match a query with the index because shapes in their definition describe the properties (RDF predicates) of the entities whereas the type index only provides the classes IRIs.
It is possible to dereference the class IRIs to get information about the properties (if available), however, it is not the current practice \cite{Taelman2023}.
A comparison of the RDF data shapes and RDF class approach due to their potential similarities is delegated to future works.~\sepfootnote{fn:litShapeComparaison}

\subsection{A Concrete Example}

We conclude this section with a concrete example.
Let's assume that a user wants to retrieve the IDs and contents of the comments in a network along with the forums ID where they have been posted and the name of the moderator of the forums.
This query is schematized in Figure~\ref{fig:shape_index}.
The query \emph{can} be represented by three star pattern queries, the comment $ Q_{comments}$, the forum $Q_{forums}$, and the moderator $Q_{moderators}$.
The full query is formed by the join of those star patterns, where the joins respect the dependencies defined by shared variables $Q = Q_{comments} \bowtie Q_{forums} \bowtie Q_{moderators}$. 
When traversing the network the query engine cannot know the content of the documents encountered, therefore, the engine \emph{must} deference every reachable document as defined by a reachability criterion.
The presence of a shape index can change the state of affairs.
If the engine encounters a domain containing exclusively book data as indicated by a complete shape index, the engine can skip the documents of the domain. 
If a domain has comment and movie review data declared by a complete shape index, the query engine can safely limit its dereferencing operations to the set of documents related to comments without affecting results completeness.
The engine can restrict its dereferencing operations because at least one star pattern is contained in the comment shape and none in the movie review and book shapes.
If the engine encounters a domain regardless of the completeness of the index, declaring comment, forum, and individual (moderators are individuals/people) data, among others, then the documents associated with the non-query-relevant part of the domain can be ignored with the same containment logic presented earlier.
Thus, we can consider that the traversal proceeds domain by domain ignoring documents known a priori to not content query-relevant data.

\section{Preliminary Results}

\sepfootnotecontent{fn:impl}{ The algorithm implementation is available at the following link \newline
\href{https://github.com/constraintAutomaton/query-shape-detection}{https://github.com/constraintAutomaton/query-shape-detection}
and the integration in the Comunica query engine at the following link 
\href{https://github.com/constraintAutomaton/comunica-feature-link-traversal/tree/feature/shapeIndex}{https://github.com/constraintAutomaton/comunica-feature-link-traversal/tree/feature/shapeIndex}.
The implementation of the benchmark and complementary results such as the analysis of the statistical significance are available at the following link 
\href{https://github.com/constraintAutomaton/amw_shape_index_results}{https://github.com/constraintAutomaton/amw\_shape\_index\_results}.
}

\sepfootnotecontent{fn:ldp}{
  \href{https://www.w3.org/TR/ldp/}{https://www.w3.org/TR/ldp/}
}

\sepfootnotecontent{fn:propertyPath}{
  \href{https://www.w3.org/TR/sparql11-query/\#propertypaths}{https://www.w3.org/TR/sparql11-query/\#propertypaths}
}

\begin{figure}[h!]
  \centering
  \includegraphics[width=\linewidth]{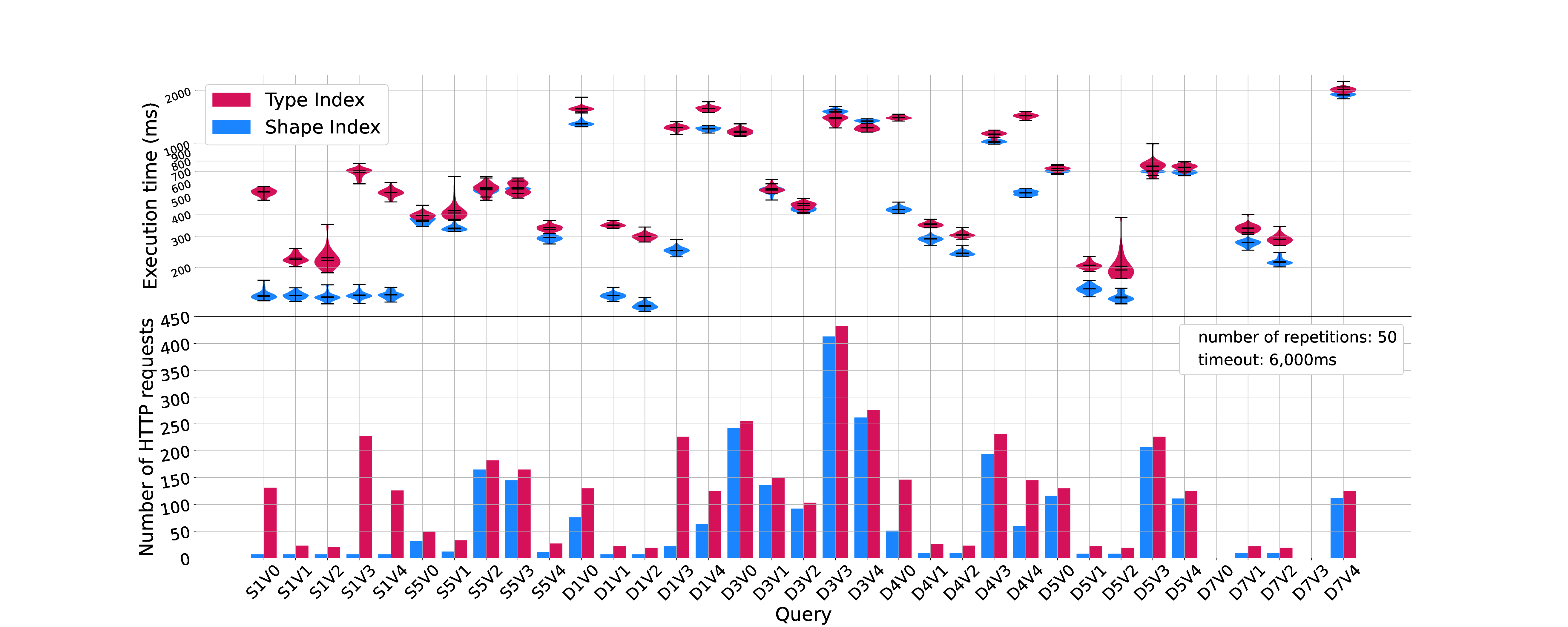}
  \caption{
  The execution time with shape indexes is consistently lower (up to 80\% with D1V3 and S1V3) or equal to that with the type indexes (except for D3V3 and D3V4), and always uses fewer HTTP requests.
  The queries are denoted with first the initial of the query template (e.g., S1 for interactive-\textbf{s}hort-\textbf{1}), and the version of the concrete query (e.g., V0). 
  Values not present in the plot (D7V0 and D7V3) indicate that the query timeout before the end of the execution.
  }
  \label{fig:result}
\end{figure}

An open-source implementation of the \href{https://github.com/constraintAutomaton/query-shape-detection}{algorithm} and an 
\href{https://github.com/constraintAutomaton/comunica-feature-link-traversal/tree/feature/shapeIndex}{integration} in the query engine 
Comunica \cite{taelman_iswc_resources_comunica_2018} is available online.~\sepfootnote{fn:impl}
We use the \href{https://github.com/SolidBench/SolidBench.js}{benchmark Solidbench} \cite{Taelman2023} to compare our approach with the current state-of-the-art (the \href{https://solid.github.io/type-indexes/}{type index} and the \href{https://www.w3.org/TR/ldp/}{LDP specification}~\sepfootnote{fn:ldp} as structural assumptions)~\cite{Taelman2023}.
We used the supported subset of SolidBench queries, skipping the currently unimplemented \href{https://www.w3.org/TR/sparql11-query/#propertypaths}{SPARQL property paths}~\sepfootnote{fn:propertyPath} and unions.
We executed each query 50 times with a timeout of 1 minute (6,000 ms).
Figure~\ref{fig:result} shows that the reduction can be as high as 80\% (D1V3 and S1V3) for execution time 
and 97\% (S1V3) for the number of HTTP requests.
Our approach reliably executes fewer HTTP requests compared to the state-of-the-art.
This is an expected result because no queries target (implicitly) each file of a user.
The shape index approach requests a subset of the request of the type index approach (without sacrificing query results) with the addition of the request to get the shape definitions which leads in general to the dereferencing of a small number of short documents.
There is not a direct correlation between the reduction of execution time and HTTP requests (e.g., the ratio 
between our approach and the state-of-the-art of the number of HTTP requests by the execution time for D1V3 is 0.5 compared to 0.15 for S1V3).
This hints at the results from the state-of-the-art \cite{Taelman2023} proposing that the query plan is the bottleneck for some queries in this environment,
however, the overhead of the containment calculation could also be a contributing factor to the current results.
In the worst cases, our approach  has similar query execution to the state-of-the-art except for D3V3 and D3V4 with an increase of 9\% of the mean of the execution time.
The variance of the execution with a shape index tends to be lower compared to the type index. 
A possible explanation for this observation is that the execution time of HTTP requests is unpredictable~\cite{hartig2016walking}
leading to an increase in variance.
This observation not only has potential implications for the reliability of multiple executions in terms of execution time
but also in terms of the performance of single executions in unstable networks where the server might take longer times to respond. 

\section{Conclusion}

\sepfootnotecontent{noChangeDataModel}{
    Considering no change in the data model.
}

\sepfootnotecontent{complexity}{
    Given the expressiveness of RDF data shapes language~\cite{delva2023, staworko_et_al:LIPIcs:2015:4985, 10.1007/978-3-319-68288-4_7} and practice in shape definitions~\cite{lieber_iswc_poster_2020, staworko_et_al:LIPIcs:2015:4985, Staworko2018ContainmentOS}.
}

The shape index approach shows that more precise source selection in LTQP can significantly reduce query execution time.
Although it is still an early effort, we believe that a solution inspired by our approach could be beneficial for the query and publication of fragmented document-based linked data.
Our solution does not require extensive computational power from the data publisher during queries and updates~\sepfootnote{noChangeDataModel} of data sources.
Additionally, using a shape index holds promise to improve the data quality of fragmented document-based linked data.
There are still multiple questions left to be answered such as how to handle private data, what is the overhead and the complexity of the method~\sepfootnote{complexity},
does the reduction of HTTP request or the reduction on the size of the internal triple store has more impact on the performance,
can the shape index alone or with other data summarisation structures be used to improve query planning without sacrificing query execution times.

\subsection*{Acknowledgements}
Supported by SolidLab Vlaanderen (Flemish Government VV023/10) and the Research Foundation Flanders (FWO) under grant number S006323N.


\bibliography{references}

\begin{thebibliography}{32}
\expandafter\ifx\csname natexlab\endcsname\relax\def\natexlab#1{#1}\fi
\providecommand{\url}[1]{\texttt{#1}}
\providecommand{\href}[2]{#2}
\providecommand{\path}[1]{#1}
\providecommand{\DOIprefix}{doi:}
\providecommand{\ArXivprefix}{arXiv:}
\providecommand{\URLprefix}{URL: }
\providecommand{\Pubmedprefix}{pmid:}
\providecommand{\doi}[1]{\href{http://dx.doi.org/#1}{\path{#1}}}
\providecommand{\Pubmed}[1]{\href{pmid:#1}{\path{#1}}}
\providecommand{\bibinfo}[2]{#2}
\ifx\xfnm\relax \def\xfnm[#1]{\unskip,\space#1}\fi
\bibitem[{Mendelzon et~al.(1996)Mendelzon, Mihaila, and Milo}]{Mendelzon1996}
\bibinfo{author}{A.~Mendelzon}, \bibinfo{author}{G.~Mihaila},
  \bibinfo{author}{T.~Milo},
\newblock \bibinfo{title}{Querying the world wide web},
\newblock in: \bibinfo{booktitle}{Fourth International Conference on Parallel
  and Distributed Information Systems}, \bibinfo{year}{1996}, pp.
  \bibinfo{pages}{80--91}. \DOIprefix\doi{10.1109/PDIS.1996.568671}.
\bibitem[{Hartig and Freytag(2012)}]{Hartig2012}
\bibinfo{author}{O.~Hartig}, \bibinfo{author}{J.-C. Freytag},
\newblock \bibinfo{title}{Foundations of traversal based query execution over
  linked data},
\newblock in: \bibinfo{booktitle}{Conference on Hypertext and Social Media}, HT
  '12, \bibinfo{publisher}{ACM}, \bibinfo{address}{New York, NY, USA},
  \bibinfo{year}{2012}, p. \bibinfo{pages}{43–52}.
  \DOIprefix\doi{10.1145/2309996.2310005}.
\bibitem[{Hartig and Özsu(2016)}]{hartig2016walking}
\bibinfo{author}{O.~Hartig}, \bibinfo{author}{M.~T. Özsu},
  \bibinfo{title}{Walking without a map: Optimizing response times of
  traversal-based linked data queries (extended version)},
  \bibinfo{year}{2016}.
\bibitem[{Hartig(2011)}]{Hartig2011}
\bibinfo{author}{O.~Hartig},
\newblock \bibinfo{title}{Zero-knowledge query planning for an iterator
  implementation of link traversal based query execution},
\newblock in: \bibinfo{booktitle}{The Semantic Web: Research and Applications},
  \bibinfo{publisher}{Springer}, \bibinfo{address}{Berlin, Heidelberg},
  \bibinfo{year}{2011}, pp. \bibinfo{pages}{154--169}.
\bibitem[{Taelman and Verborgh(2023)}]{Taelman2023}
\bibinfo{author}{R.~Taelman}, \bibinfo{author}{R.~Verborgh},
\newblock \bibinfo{title}{Link traversal query processing over decentralized
  environments with structural assumptions},
\newblock in: \bibinfo{booktitle}{Proceedings of the 22nd International
  Semantic Web Conference}, \bibinfo{year}{2023}.
\bibitem[{Verborgh and Taelman(2020)}]{verborgh2020guided}
\bibinfo{author}{R.~Verborgh}, \bibinfo{author}{R.~Taelman},
  \bibinfo{title}{Guided link-traversal-based query processing},
  \bibinfo{year}{2020}. \href{http://arxiv.org/abs/2005.02239}{{\tt
  arXiv:2005.02239}}.
\bibitem[{Fielding(2000)}]{Fielding}
\bibinfo{author}{R.~T. Fielding}, \bibinfo{title}{Architectural styles and the
  design of network-based doftware architectures}, Ph.D. thesis, University of
  California, \bibinfo{year}{2000}.
\bibitem[{Taelman and Verborgh(2017)}]{Taelman2017}
\bibinfo{author}{R.~Taelman}, \bibinfo{author}{R.~Verborgh},
\newblock \bibinfo{title}{Declaratively describing responses of
  hypermedia-driven web apis},
\newblock in: \bibinfo{booktitle}{Knowledge Capture Conference}, K-CAP '17,
  \bibinfo{publisher}{Association for Computing Machinery},
  \bibinfo{address}{New York, NY, USA}, \bibinfo{year}{2017}.
  \DOIprefix\doi{10.1145/3148011.3154467}.
\bibitem[{Eschauzier et~al.(2023)Eschauzier, Taelman, and
  Verborgh}]{eschauzier_quweda_2023}
\bibinfo{author}{R.~Eschauzier}, \bibinfo{author}{R.~Taelman},
  \bibinfo{author}{R.~Verborgh},
\newblock \bibinfo{title}{How does the link queue evolve during traversal-based
  query processing?},
\newblock in: \bibinfo{booktitle}{Proceedings of the 7th QuWeDa}, CEUR Workshop
  Proceedings, \bibinfo{year}{2023}.
\bibitem[{Gayo et~al.(2018)Gayo, Prud'hommeaux, Boneva, and
  Kontokostas}]{Gayo2018a}
\bibinfo{author}{J.-E.~L. Gayo}, \bibinfo{author}{E.~Prud'hommeaux},
  \bibinfo{author}{I.~Boneva}, \bibinfo{author}{D.~Kontokostas},
  \bibinfo{title}{Validating RDF Data: Applications},
  \bibinfo{publisher}{Springer International Publishing},
  \bibinfo{address}{Cham}, \bibinfo{year}{2018}, pp. \bibinfo{pages}{195--231}.
  \DOIprefix\doi{10.1007/978-3-031-79478-0_6}.
\bibitem[{Rabbani et~al.(2021)Rabbani, Lissandrini, and Hose}]{kashif2021}
\bibinfo{author}{K.~Rabbani}, \bibinfo{author}{M.~Lissandrini},
  \bibinfo{author}{K.~Hose}, \bibinfo{title}{Optimizing sparql queries using
  shape statistics}, \bibinfo{year}{2021}.
  \DOIprefix\doi{10.5441/002/EDBT.2021.59}.
\bibitem[{Stuckenschmidt et~al.(2004)Stuckenschmidt, Vdovjak, Houben, and
  Broekstra}]{Stuckenschmidt2004}
\bibinfo{author}{H.~Stuckenschmidt}, \bibinfo{author}{R.~Vdovjak},
  \bibinfo{author}{G.-J. Houben}, \bibinfo{author}{J.~Broekstra},
\newblock \bibinfo{title}{Index structures and algorithms for querying
  distributed rdf repositories},
\newblock in: \bibinfo{booktitle}{Proceedings of the 13th International
  Conference on World Wide Web}, WWW '04, \bibinfo{publisher}{Association for
  Computing Machinery}, \bibinfo{address}{New York, NY, USA},
  \bibinfo{year}{2004}, p. \bibinfo{pages}{631–639}. \URLprefix
  \url{https://doi.org/10.1145/988672.988758}.
  \DOIprefix\doi{10.1145/988672.988758}.
\bibitem[{Goldman and Widom(1997)}]{Goldman1997}
\bibinfo{author}{R.~Goldman}, \bibinfo{author}{J.~Widom},
\newblock \bibinfo{title}{Dataguides: Enabling query formulation and
  optimization in semistructured databases},
\newblock in: \bibinfo{booktitle}{Proceedings of the 23rd International
  Conference on Very Large Data Bases}, VLDB '97, \bibinfo{publisher}{Morgan
  Kaufmann Publishers Inc.}, \bibinfo{address}{San Francisco, CA, USA},
  \bibinfo{year}{1997}, p. \bibinfo{pages}{436–445}.
\bibitem[{Harth et~al.(2010)Harth, Hose, Karnstedt, Polleres, Sattler, and
  Umbrich}]{Harth2010}
\bibinfo{author}{A.~Harth}, \bibinfo{author}{K.~Hose},
  \bibinfo{author}{M.~Karnstedt}, \bibinfo{author}{A.~Polleres},
  \bibinfo{author}{K.-U. Sattler}, \bibinfo{author}{J.~Umbrich},
\newblock \bibinfo{title}{Data summaries for on-demand queries over linked
  data},
\newblock in: \bibinfo{booktitle}{Proceedings of the 19th International
  Conference on World Wide Web}, WWW '10, \bibinfo{publisher}{Association for
  Computing Machinery}, \bibinfo{address}{New York, NY, USA},
  \bibinfo{year}{2010}, p. \bibinfo{pages}{411–420}. \URLprefix
  \url{https://doi.org/10.1145/1772690.1772733}.
  \DOIprefix\doi{10.1145/1772690.1772733}.
\bibitem[{Neumann and Moerkotte(2011)}]{Neumann2011CharacteristicSA}
\bibinfo{author}{T.~Neumann}, \bibinfo{author}{G.~Moerkotte},
\newblock \bibinfo{title}{Characteristic sets: Accurate cardinality estimation
  for rdf queries with multiple joins},
\newblock \bibinfo{journal}{2011 IEEE 27th International Conference on Data
  Engineering}  (\bibinfo{year}{2011}) \bibinfo{pages}{984--994}.
\bibitem[{Meimaris and Papastefanatos(2018)}]{Meimaris2018HierarchicalCS}
\bibinfo{author}{M.~Meimaris}, \bibinfo{author}{G.~Papastefanatos},
\newblock \bibinfo{title}{Hierarchical characteristic set merging for
  optimizing sparql queries in heterogeneous rdf},
\newblock \bibinfo{journal}{ArXiv} \bibinfo{volume}{abs/1809.02345}
  (\bibinfo{year}{2018}).
\bibitem[{Gayo et~al.(2018{\natexlab{a}})Gayo, Prud'hommeaux, Boneva, and
  Kontokostas}]{Gayo2018}
\bibinfo{author}{J.-E.~L. Gayo}, \bibinfo{author}{E.~Prud'hommeaux},
  \bibinfo{author}{I.~Boneva}, \bibinfo{author}{D.~Kontokostas},
  \bibinfo{title}{Shape Expressions}, \bibinfo{publisher}{Springer
  International Publishing}, \bibinfo{address}{Cham},
  \bibinfo{year}{2018}{\natexlab{a}}, pp. \bibinfo{pages}{55--117}.
  \DOIprefix\doi{10.1007/978-3-031-79478-0_4}.
\bibitem[{Gayo et~al.(2018{\natexlab{b}})Gayo, Prud'hommeaux, Boneva, and
  Kontokostas}]{Gayo2018Shacl}
\bibinfo{author}{J.-E.~L. Gayo}, \bibinfo{author}{E.~Prud'hommeaux},
  \bibinfo{author}{I.~Boneva}, \bibinfo{author}{D.~Kontokostas},
  \bibinfo{title}{SHACL}, \bibinfo{publisher}{Springer International
  Publishing}, \bibinfo{address}{Cham}, \bibinfo{year}{2018}{\natexlab{b}}, pp.
  \bibinfo{pages}{119--194}. \URLprefix
  \url{https://doi.org/10.1007/978-3-031-79478-0_5}.
  \DOIprefix\doi{10.1007/978-3-031-79478-0_5}.
\bibitem[{Foto~Afrati(2019)}]{afariQCE}
\bibinfo{author}{R.~C. Foto~Afrati}, \bibinfo{title}{Query Containment and
  Equivalence}, \bibinfo{publisher}{Springer Cham}, \bibinfo{year}{2019}, pp.
  \bibinfo{pages}{21--59}.
  \DOIprefix\doi{https://.doi.org/10.1007/978-3-031-01871-8}.
\bibitem[{Spasi{\'{c}} and Jani{\v{c}}i{\'{c}}(2023)}]{Spasi2023}
\bibinfo{author}{M.~Spasi{\'{c}}}, \bibinfo{author}{M.~V. Jani{\v{c}}i{\'{c}}},
\newblock \bibinfo{title}{Solving the {SPARQL} query containment problem with
  {SpeCS}},
\newblock \bibinfo{journal}{Journal of Web Semantics} \bibinfo{volume}{76}
  (\bibinfo{year}{2023}) \bibinfo{pages}{100770}.
  \DOIprefix\doi{10.1016/j.websem.2022.100770}.
\bibitem[{Chekol et~al.(2018)Chekol, Euzenat, Genevès, and
  Layaïda}]{Chekol2018}
\bibinfo{author}{M.~W. Chekol}, \bibinfo{author}{J.~Euzenat},
  \bibinfo{author}{P.~Genevès}, \bibinfo{author}{N.~Layaïda},
\newblock \bibinfo{title}{Sparql query containment under schema},
\newblock \bibinfo{journal}{Journal on Data Semantics} \bibinfo{volume}{7}
  (\bibinfo{year}{2018}) \bibinfo{pages}{133–154}. \URLprefix
  \url{http://dx.doi.org/10.1007/s13740-018-0087-1}.
  \DOIprefix\doi{10.1007/s13740-018-0087-1}.
\bibitem[{{Delva, Thomas and Dimou, Anastasia and Jakubowksi, Maxime and Van
  den Bussche, Jan}(2023)}]{delva2023}
\bibinfo{author}{{Delva, Thomas and Dimou, Anastasia and Jakubowksi, Maxime and
  Van den Bussche, Jan}},
\newblock \bibinfo{title}{{Data provenance for SHACL}},
\newblock in: \bibinfo{booktitle}{{Proceedings 26th International Conference on
  Extending Database Technology (EDBT 2023)}}, volume~\bibinfo{volume}{{26}},
  \bibinfo{year}{{2023}}, pp. \bibinfo{pages}{{285--297}}. \URLprefix
  \url{{http://doi.org/10.48786/edbt.2023.23}}.
\bibitem[{W3C(2013)}]{spapeExpressionConvert}
\bibinfo{author}{W3C}, \bibinfo{title}{Sparql queries to validate shape
  expressions (informative)}, \bibinfo{year}{2013}. \URLprefix
  \url{https://www.w3.org/2013/ShEx/toSPARQL.html}.
\bibitem[{Gayo et~al.(2017)Gayo, Prud'hommeaux, Solbrig, and
  Boneva}]{labragayo2017validating}
\bibinfo{author}{J.-E.~L. Gayo}, \bibinfo{author}{E.~Prud'hommeaux},
  \bibinfo{author}{H.~Solbrig}, \bibinfo{author}{I.~Boneva},
  \bibinfo{title}{Validating and describing linked data portals using shapes},
  \bibinfo{year}{2017}. \href{http://arxiv.org/abs/1701.08924}{{\tt
  arXiv:1701.08924}}.
\bibitem[{Corman et~al.(2019)Corman, Florenzano, Reutter, and
  Savkovi{\'{c}}}]{Corman2019}
\bibinfo{author}{J.~Corman}, \bibinfo{author}{F.~Florenzano},
  \bibinfo{author}{J.~L. Reutter}, \bibinfo{author}{O.~Savkovi{\'{c}}},
\newblock \bibinfo{title}{Validating shacl constraints over a sparql endpoint},
\newblock in: \bibinfo{booktitle}{The Semantic Web -- ISWC 2019},
  \bibinfo{publisher}{Springer International Publishing},
  \bibinfo{address}{Cham}, \bibinfo{year}{2019}, pp. \bibinfo{pages}{145--163}.
\bibitem[{Yang et~al.(2021)Yang, Youill, Woicik, Liu, Yu, Serafini, Aboulnaga,
  and Stonebraker}]{Yang2021FlexPushdownDBHP}
\bibinfo{author}{Y.~Yang}, \bibinfo{author}{M.~Youill},
  \bibinfo{author}{M.~Woicik}, \bibinfo{author}{Y.~Liu},
  \bibinfo{author}{X.~Yu}, \bibinfo{author}{M.~Serafini},
  \bibinfo{author}{A.~Aboulnaga}, \bibinfo{author}{M.~Stonebraker},
\newblock \bibinfo{title}{Flexpushdowndb: Hybrid pushdown and caching in a
  cloud dbms},
\newblock \bibinfo{journal}{Proc. VLDB Endow.} \bibinfo{volume}{14}
  (\bibinfo{year}{2021}) \bibinfo{pages}{2101--2113}.
\bibitem[{De~Meester et~al.(2021)De~Meester, Heyvaert, Arndt, Dimou, and
  Verborgh}]{demeester_swj_2021}
\bibinfo{author}{B.~De~Meester}, \bibinfo{author}{P.~Heyvaert},
  \bibinfo{author}{D.~Arndt}, \bibinfo{author}{A.~Dimou},
  \bibinfo{author}{R.~Verborgh},
\newblock \bibinfo{title}{{RDF} graph validation using rule-based reasoning},
\newblock \bibinfo{journal}{Semantic Web Journal} \bibinfo{volume}{12}
  (\bibinfo{year}{2021}) \bibinfo{pages}{117--142}.
  \DOIprefix\doi{10.3233/SW-200384}.
\bibitem[{Taelman et~al.(2018)Taelman, Van~Herwegen, Vander~Sande, and
  Verborgh}]{taelman_iswc_resources_comunica_2018}
\bibinfo{author}{R.~Taelman}, \bibinfo{author}{J.~Van~Herwegen},
  \bibinfo{author}{M.~Vander~Sande}, \bibinfo{author}{R.~Verborgh},
\newblock \bibinfo{title}{Comunica: a modular sparql query engine for the web},
\newblock in: \bibinfo{booktitle}{Proceedings of the 17th International
  Semantic Web Conference}, \bibinfo{year}{2018}.
\bibitem[{Staworko et~al.(2015)Staworko, Boneva, Gayo, Hym, Prud'hommeaux, and
  Solbrig}]{staworko_et_al:LIPIcs:2015:4985}
\bibinfo{author}{S.~Staworko}, \bibinfo{author}{I.~Boneva},
  \bibinfo{author}{J.-E.~L. Gayo}, \bibinfo{author}{S.~Hym},
  \bibinfo{author}{E.~G. Prud'hommeaux}, \bibinfo{author}{H.~Solbrig},
\newblock \bibinfo{title}{{Complexity and Expressiveness of ShEx for RDF}},
\newblock in: \bibinfo{booktitle}{18th International Conference on Database
  Theory (ICDT 2015)}, volume~\bibinfo{volume}{31} of
  \textit{\bibinfo{series}{Leibniz International Proceedings in Informatics
  (LIPIcs)}}, \bibinfo{publisher}{Schloss Dagstuhl--Leibniz-Zentrum fuer
  Informatik}, \bibinfo{address}{Dagstuhl, Germany}, \bibinfo{year}{2015}, pp.
  \bibinfo{pages}{195--211}. \DOIprefix\doi{10.4230/LIPIcs.ICDT.2015.195}.
\bibitem[{Boneva et~al.(2017)Boneva, Gayo, and
  Prud’hommeaux}]{10.1007/978-3-319-68288-4_7}
\bibinfo{author}{I.~Boneva}, \bibinfo{author}{J.-E.~L. Gayo},
  \bibinfo{author}{E.~G. Prud’hommeaux},
\newblock \bibinfo{title}{Semantics and validation of shapes schemas for rdf},
\newblock in: \bibinfo{booktitle}{The Semantic Web – ISWC 2017: 16th
  International Semantic Web Conference, Vienna, Austria, October 21–25,
  2017, Proceedings, Part I}, \bibinfo{publisher}{Springer-Verlag},
  \bibinfo{address}{Berlin, Heidelberg}, \bibinfo{year}{2017}, p.
  \bibinfo{pages}{104–120}. \DOIprefix\doi{10.1007/978-3-319-68288-4_7}.
\bibitem[{Lieber et~al.(2020)Lieber, Dimou, and
  Verborgh}]{lieber_iswc_poster_2020}
\bibinfo{author}{S.~Lieber}, \bibinfo{author}{A.~Dimou},
  \bibinfo{author}{R.~Verborgh},
\newblock \bibinfo{title}{Statistics about data shape use in {RDF} data},
\newblock in: \bibinfo{booktitle}{Proceedings of the 19th International
  Semantic Web Conference: Posters, Demos, and Industry Tracks}, volume
  \bibinfo{volume}{2721} of \textit{\bibinfo{series}{CEUR Workshop
  Proceedings}}, \bibinfo{year}{2020}, pp. \bibinfo{pages}{330--335}.
  \URLprefix \url{http://ceur-ws.org/Vol-2721/paper584.pdf}.
\bibitem[{Staworko and Wieczorek(2018)}]{Staworko2018ContainmentOS}
\bibinfo{author}{S.~Staworko}, \bibinfo{author}{P.~Wieczorek},
\newblock \bibinfo{title}{Containment of shape expression schemas for rdf},
\newblock \bibinfo{journal}{Proceedings of the 38th ACM SIGMOD-SIGACT-SIGAI
  Symposium on Principles of Database Systems}  (\bibinfo{year}{2018}).

\end{thebibliography}

\end{document}